# EXPERT-LEVEL ANNOTATION QUALITY ACHIEVED BY GAMIFIED CROWDSOURCING FOR B-LINE SEGMENTATION IN LUNG ULTRASOUND

*Mike Jin[1,2], Nicole M Duggan[2], Varoon Bashyakarla[1], Maria Alejandra Duran Mendicuti[2], Stephen Hallisey[2], Denie Bernier[2], Joseph Stegeman[3], Erik Duhaime[1], Tina Kapur[2], Andrew J Goldsmith[2]*

[1]Centaur Labs, Boston, MA, [2]Brigham and Women's Hospital, Harvard Medical School, Boston, MA, [3]Highland Hospital, Alameda Health System, Oakland, USA

## ABSTRACT

Accurate and scalable annotation of medical data is critical for the development of medical AI, but obtaining time for annotation from medical experts is challenging. Gamified crowdsourcing has demonstrated potential for obtaining highly accurate annotations for medical data at scale, and we demonstrate the same in this study for the segmentation of B-lines, an indicator of pulmonary congestion, on still frames within point-of-care lung ultrasound clips. We collected 21,154 annotations from 214 annotators over 2.5 days, and we demonstrated that the concordance of crowd consensus segmentations with reference standards exceeds that of individual experts with the same reference standards, both in terms of B-line count (mean squared error 0.239 vs. 0.308, $p<0.05$) as well as the spatial precision of B-line annotations (mean Dice-H score 0.755 vs. 0.643, $p<0.05$). These results suggest that expert-quality segmentations can be achieved using gamified crowdsourcing.

## 1. INTRODUCTION

### 1.1. Need for high-quality medical AI/ML annotations

Artificial intelligence and machine learning (AI/ML) demonstrate great promise for increasing the accuracy of clinical decision-making, lightening the onus on healthcare practitioners, and improving the efficiency of diagnostic workflows, especially in the context of medical imaging [1]. These technologies inform and streamline medical decision-making by assisting the analysis of clinical images. Since these interventions can be used directly at the point of care, they may save time and reduce errors. Such technologies rely on high-quality labeled imaging data, which can be challenging and cost prohibitive to reliably collect, since data can be unstructured and since enlisting medical expert annotators to annotate large datasets tends to be expensive and ill-suited to scaling.

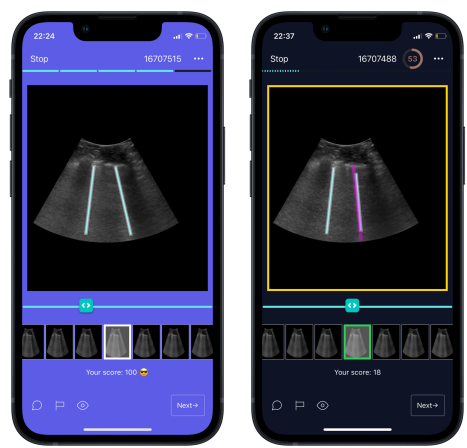

**Figure 1. a)** DiagnosUs user interface for a feedback case with two B-lines. The user-annotated B-lines coincide with the known B-lines. **b)** A feedback case in which the user correctly identified one B-line but misplaced the second B-line. Correct answers (magenta) are shown as feedback.

### 1.2. Lung POCUS B-line segmentation

Point-of-care ultrasound (POCUS) is a highly valuable clinical tool that can be used at patients' bedside to make important diagnoses, especially in emergency medicine [2-4]. Thus, there is substantial clinical interest in the development of AI models which can facilitate and ensure greater consistency in lung ultrasound (LUS) interpretation [5-10].

B-lines are linear artifacts that are known to be markers of pulmonary congestion and can indicate severity of many conditions including inflammatory lung disease, pneumonia, and congestive heart failure exacerbations [3, 5, 11]. B-lines materialize in LUS in sync with the respiratory cycle, and ML models have been successful in identifying and quantifying B-lines present in LUS [6 - 10, 12]. Crowdsourced classification of B-lines has previously been shown to achieve expert-level quality [13]. This study is the first to apply gamified crowdsourcing to LUS B-line segmentation.

### 1.3. Leveraging gamification and collective intelligence for scalable, accurate annotation acquisition

Intelligently aggregating opinions from multiple annotators to leverage the "wisdom of the crowd" has been shown to produce superior outcomes for a variety of tasks, including data annotation and forecasting. Soliciting many crowdsourced opinions can thus enable scalable medical data annotation, and several studies across a range of medical settings have demonstrated that large numbers of non-experts can rival and even outperform individual experts with respect to the accuracy of their annotations when interpreting medical image data while reducing costs and increasing efficiency [13 - 18].

Generating trustworthy annotations via crowdsourcing requires robust quality control and incentive management. DiagnosUs – a free iOS application on which thousands of participants compete daily for cash prizes on a variety of medical data annotation tasks – achieves this by evaluating annotators on their submitted opinions and providing feedback to reinforce their skills [18]. Each user's annotation accuracy on randomly interspersed cases having ground truth annotations is revealed to them as feedback after their opinion submission (Figure 1), and is additionally used both to determine their leaderboard score for prize eligibility and to estimate the user's skill. The opinions from the most skilled users can then be combined to form crowd consensus opinions that previous studies have shown to be highly accurate on several classification tasks [13,18].

### 1.4. Objectives

Our goal was to assess whether crowdsourced consensus B-line segmentations could achieve concordance with reference standards comparable to that of individual experts. We assessed annotation concordance based on B-line count as well as based on spatial deviation. If crowd consensus segmentations prove to be comparable to those of experts, then properly incentivized gamified crowdsourcing could be a viable alternative to gathering expert-quality annotations in other medical contexts.

## 2. METHODS

### 2.1. Data procurement and dataset construction

All lung POCUS data from this study were acquired from 203 patients admitted to the emergency department of Brigham and Women's Hospital with shortness of breath between March 1, 2020, and February 28, 2022. A previous study [13] split these 203 patients 50/50 into a training (102 patients) and a test set (101 patients), classifying 393 POCUS video clips by a majority-vote consensus of six lung ultrasound experts as having (141 clips) or not having (252 clips) B-lines anywhere within the clip. 81 of the 141 clips with B-lines belonged to the training set of patients, and 200 single frames for crowd training were randomly sampled from these clips. Likewise, 200 holdout frames for crowd evaluation were sampled from the remaining 60 clips with B-lines. Reference standard segmentations were created for all 400 sampled frames based on a consensus (see section 2.4) of segmentations from five experts (Figure 2).

Expert and crowdsourced opinions were all collected using DiagnosUs (Centaur Labs, Boston, USA). All line coordinates were scaled to [0,100] relative to the dimensions of the image on which they were annotated.

### 2.2. Segmentation task definition

Five lung ultrasound experts and all crowd participants received instructions to draw all B-lines present on single POCUS frames with the 10 previous and 10 next frames viewable for context. The instructions also specified that: A) endpoints should be placed at the pleural line and the bottom of the sonographic field, B) confluent B-lines should be annotated using just a single line down the middle of the confluent B-line, and C) B-lines originating from the diaphragm or consolidated lung as well as comet tail (i.e. hyperechoic lines that do not extend to the bottom of the sonographic field) should not be annotated. For crowd annotators, completion of these instructions was required for participating in the contest.

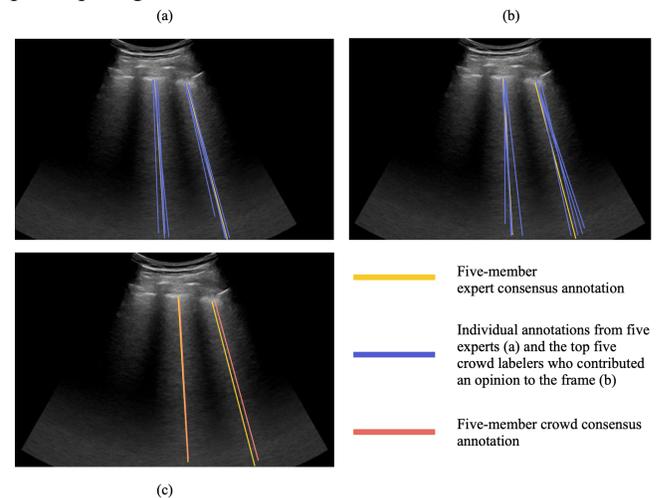

**Figure 2**. Individual and consensus annotations showing variation in annotations from experts and top crowd members for a single POCUS frame.

### 2.3. Dice-H scores for B-line annotation similarity

In addition to comparing annotations for similarity simply based on the number of B-lines, to account for positional similarity we used **Dice-H score**, a Dice score-inspired metric using Hausdorff distance. To calculate Dice-H for two B-line annotations (i.e., two sets of lines), we first found a bijection between the two sets using the Hungarian algorithm (Jonker-Volgenant variant) to get a matching with maximum total similarity, using a similarity score for line pairings that is at most 1 for a Hausdorff distance of 0 and decays linearly to a minimum of 0 for a Hausdorff distance of 5 or more. Note this meant some B-lines from the larger set would be unmatched if the two sets contained different numbers of B-lines. We then calculated Dice-H as two times the sum of the similarity scores across all the matched pairs, divided by the total number of B-lines in the two annotations. Dice-H was used to score in-game user annotations as well, but with a linearly decaying score to 0 at a Hausdorff distance of 10 rather than 5 to minimize user discouragement.

### 2.4. Determining consensus annotations via clustering

To form a consensus annotation from multiple annotators' opinions (Figure 2), all B-lines from all annotators were clustered by agglomerative clustering up to a Hausdorff distance of 10. If multiple B-lines from the same annotator were assigned to the same terminal cluster, only the B-line closest to the cluster centroid was kept. Then, clusters containing a B-line count less than or equal to 50% of the number of original annotators were discarded. Finally, all remaining clusters produced a consensus annotation B-line by averaging the lines within the cluster.

### 2.5. Reference standard annotations and concordance

For crowd training, the "ground truth" annotations on the 200 training frames were constructed based on a consensus of all five experts. For evaluation of crowd consensus annotations on the 200 test frames, the concordance between crowd consensus annotations and reference standards was compared to the concordance between individual expert annotations with the same reference standards. Each individual expert annotation was compared to a leave-one-out consensus of the other four experts, and crowd consensus annotations were compared to the same five leave-one-out expert consensus annotations. Both crowd and expert concordance figures reported are therefore an average over five concordances, one for each leave-one-out consensus annotation. For evaluation of concordance based on B-line counts, reference standard counts were computed by the average count of the contributing experts.

### 2.6. Gamified crowdsourced annotation

Crowd annotators were shown the 400 training and test frames in random order with a relative frequency of 1:2, with training frames having B-lines shown equally frequently as training frames having no B-lines, and with test frames sampled uniformly. Annotators were shown the reference standard annotation (formed by a consensus of all five experts) and their Dice-H score as feedback following their submission of an opinion on a training frame, but received no feedback on test frames. Since users' Dice-H scores determined their leaderboard score and thus their eligibility for cash prizes (Figure 1), and because they did not know *a priori* which frames were training frames, they were incentivized to give maximal effort toward all annotations, such that their performance on training cases (i.e., trailing average Dice-H score) could be used as a proxy for their skill at any given point in time. Each opinion was thus associated with a **Qscore** (i.e. Quality score), the user's trailing average Dice-H score at the time they submitted the opinion. Crowd consensus annotations were constructed using only the five opinions (mirroring the number of experts involved in this study) with the highest Qscores on each test case, and only the most recent opinion per user per case.

### 3. RESULTS

A total of 21,154 opinions were collected from 214 users over two contests spanning 60 hours, reflecting a mean acquisition rate of 5.9 opinions per minute and 98.9 opinions per user. For the 200 test cases, 13,939 opinions were collected, of which 1,000 opinions (from users with the top five Qscores per case) belonging to 46 users were used to generate the final crowd consensus annotations. The total prize pool per contest was $100 (USD), awarded to the top annotators who annotated at least 75 cases. The training and evaluation data were comparable with respect to the number of B-lines present in each clip.

To assess the concordance of the crowd consensus with reference standards relative to that of individual experts, we computed the mean squared error (MSE) for the number of B-lines annotated (Figure 3a). Reference standards were constructed as the average of B-line counts from experts (see section 2.5). The crowd consensus outperformed the average individual expert in estimating the number of B-lines present (0.239 vs. 0.308, $p<0.05$, Student's *t*).

To quantify spatial similarity in B-line annotations, we measured the mean Dice-H score of individual expert annotations relative to reference standards (0.643), and compared that to the mean Dice-H score of the crowd consensus annotations with the same reference standards (0.755). This discrepancy again reflected an outperformance in concordance, this time in annotation precision, of the crowd consensus relative to individual experts ($p<0.05$, bias-corrected and accelerated 95% bootstrap CI (0.071, 0.148) for the difference in Dice-H score between crowd and expert with respect to the same reference standards). The accelerated bootstrapped distribution for the difference in Dice-H scores between the crowd consensus and expert consensus is shown in Figure 3b.

Increased variation of expert opinions for a case can be a measure of its ambiguity, and we find that cases with increased expert opinion variation also tend to be cases with increased variation in the top crowd opinions. When we define the expert agreement on a given case as the average of all pairwise Dice-H scores of the five experts, and likewise for crowd agreement using the five crowd annotations with the highest Qscore, the correlation between crowd and expert agreement across all test cases was significant (Pearson's $r$=0.691, p<0.001). Similarly, Dice-H scores between the crowd consensus and reference standards also tended to be correlated with Dice-H scores between individual expert opinions and the same reference standards (Pearson's $r$=0.657, p<0.001). These results suggest that when the crowd consensus deviated more from the reference standards, usually the individual expert annotations were somewhat in disagreement as well. For example, when the crowd consensus annotation differed in B-line count from the five-expert reference standard, 85% of the time there was at least one expert opinion which matched the crowd consensus in B-line counts.

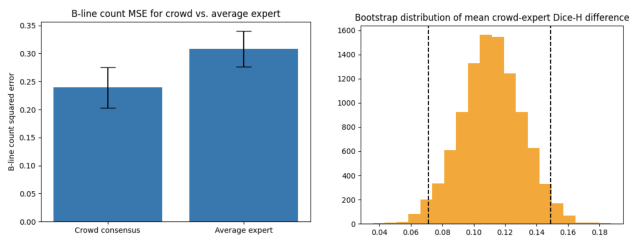

**Figure 3. a)** Comparison of B-line count MSE between crowd consensus annotations and the average expert. **b)** The bias-corrected, accelerated bootstrap distribution for the mean difference in Dice-H scores between the crowd and the expert consensus, both measured with respect to the same reference standards.

Finally, since medical learning and knowledge acquisition is a strong motivating factor for many DiagnosUs users, we generated a learning curve to examine trends in crowd annotation performance (Figure 4). The plot shows that, as the number of cases seen by a given user increased, the concordance of that user's annotations with reference standards also tended to increase, demonstrating that user performance improved with time with feedback, nearly reaching the average Dice-H of individual experts (dotted line). The dashed line shows the "wisdom of the crowd" effect: by combining the top 5 crowd annotations on each case, the resulting Dice-H scores of crowd consensus annotations with respect to reference standards reach a level exceeding that of individual experts, as previously discussed (cf. Figure 3b).

## 4. DISCUSSION

The development of medical AI/ML models for assessing medical imaging to improve clinical outcomes, streamline medical decision-making, and reduce costs relies on the availability of trusted annotations. Determining high-quality medical annotations from crowdsourced opinions presents an effective, scalable, and cost-effective alternative to traditional means of collecting this information from medical experts. We demonstrate that intelligent aggregation of crowd opinions yields results that are comparable to that of experts with respect to both the number of B-lines and to the spatial precision of where B-lines are located in lung POCUS.

These findings contribute to the literature indicating that large groups can collectively match and even surpass the performance of specialists through leveraging the "wisdom of the crowd" [19]. To-date, gamified crowdsourcing has been shown to produce results on par with experts in tasks related to skin lesion classification and B-line lung POCUS [13, 18]. This study is the first we know of to illustrate the benefits of gamified crowdsourcing on B-line lung POCUS tasks. Given this promising and growing body of evidence, it is possible that the benefits of gamified crowdsourcing may extend to other medical contexts as well.

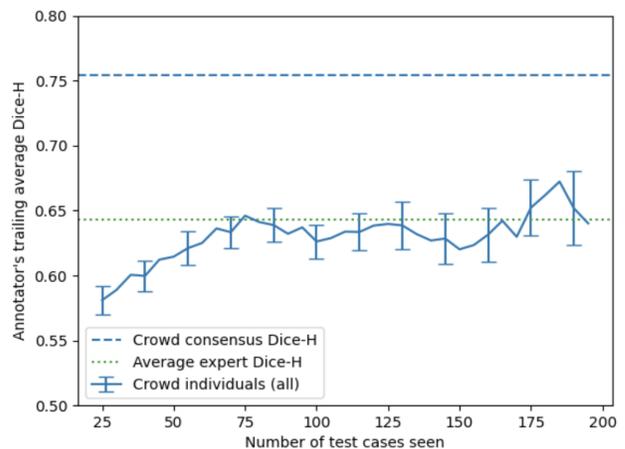

**Figure 4.** Learning curve for crowd users, suggesting skill improvement as measured by the trailing average Dice-H of the user's annotations with respect to reference standards. Error bars represent standard error of the mean.

## 5. COMPLIANCE WITH ETHICAL STANDARDS

Centaur Labs is a HIPAA-compliant organization and all data was handled in compliance with HIPAA privacy and security standards. Crowdsourced annotations obtained from individual users through DiagnosUs did not require separate ethical approval due to the DiagnosUs User Agreement. The DiagnosUs app is free to use and all annotators opted into the study and agreed that their data could be used for research purposes and be analyzed through the DiagnosUs User Agreement.

## 6. ACKNOWLEDGMENTS

Philips and Exo. This investigation was supported by a Massachusetts Life Sciences Center (MLSC) grant.## 7. REFERENCES

[1] S. S. Han et al., "Augmented Intelligence Dermatology: Deep Neural Networks Empower Medical Professionals in Diagnosing Skin Cancer and Predicting Treatment Options for 134 Skin Disorders," J Invest Dermatol, vol. 140, no. 9, pp. 1753–1761, Sep. 2020, doi: 10.1016/j.jid.2020.01.019.

[2] Lichtenstein DA. BLUE-protocol and FALLS-protocol: two applications of lung ultrasound in the critically ill. Chest. 2015 Jun;147(6):1659-1670. doi: 10.1378/chest.14-1313.

[3] ACEP Policy Statement. Ultrasound Guidelines: Emergency, Point-of-Care and Clinical Ultrasound Guidelines in Medicine. Ann Emerg Med. 2017 May;69(5):e27-e54. doi: 10.1016/j.annemergmed.

[4] Blehar DJ, Barton B, Gaspari RJ. Learning curves in emergency ultrasound education. Acad Emerg Med. 2015 May;22(5):574-82. doi: 10.1111/acem.12653.

[5] Arntfield R, Wu D, Tschirhart J, et al. Automation of Lung Ultrasound Interpretation via Deep Learning for the Classification of Normal versus Abnormal Lung Parenchyma: A Multicenter Study. Diagnostics (Basel). 2021 Nov 4;11(11):2049. doi: 10.3390/diagnostics11112049.

[6] Baloescu C, Toporek G, Kim S, et al. Automated Lung Ultrasound B-Line Assessment Using a Deep Learning Algorithm. IEEE Trans Ultrason Ferroelectr Freq Control. 2020 Nov;67(11):2312-2320. doi: 10.1109/TUFFC.2020.3002249.

[7] Moore CL, Wang J, Battisti AJ, et al. Interobserver Agreement and Correlation of an Automated Algorithm for B-Line Identification and Quantification With Expert Sonologist Review in a Handheld Ultrasound Device. J Ultrasound Med. 2022 Oct;41(10):2487-2495. doi: 10.1002/jum.15935.

[8] Pare JR, Gjesteby LA, Telfer BA, et al. Transfer Learning for Automated COVID-19 B-Line Classification in Lung Ultrasound. Annu Int Conf IEEE Eng Med Biol Soc. 2022 Jul;2022:1675-1681. doi: 10.1109/EMBC48229.2022.9871894.

[9] van Sloun RJG, Demi L. Localizing B-Lines in Lung Ultrasonography by Weakly Supervised Deep Learning, In-Vivo Results. IEEE J Biomed Health Inform. 2020 Apr;24(4):957-964. doi: 10.1109/JBHI.2019.2936151.

[10] Wang J, Yang X, Zhou B, et al. Review of Machine Learning in Lung Ultrasound in COVID- 19 Pandemic. J Imaging. 2022 Mar 5;8(3):65. doi: 10.3390/jimaging8030065.

[11] Pang PS, Russell FM, Ehrman R, et al. Lung Ultrasound-Guided Emergency Department Management of Acute Heart Failure (BLUSHED-AHF): A Randomized Controlled Pilot Trial. JACC Heart Fail. 2021 Sep;9(9):638-648. doi: 10.1016/j.jchf.2021.05.008.

[12] Horos Project. https://horosproject.org/ (2018).

[13] N. Duggan et al., "Gamified Crowdsourcing as a Novel Approach to Lung Ultrasound Dataset Labeling," arXiv (Cornell University), Jun. 2023, doi: https://doi.org/10.48550/arxiv.2306.06773.

[14] P. Tschandl et al., "Comparison of the accuracy of human readers versus machine-learning algorithms for pigmented skin lesion classification: an open, web-based, international, diagnostic study," Lancet Oncol, vol. 20, no. 7, pp. 938– 947, Jul. 2019, doi: 10.1016/S1470-2045(19)30333-X.

[15] P. Rajpurkar et al., "CheXNet: Radiologist-Level Pneumonia Detection on Chest X-Rays with Deep Learning." arXiv, Dec. 25, 2017. doi: 10.48550/arXiv.1711.05225.

[16] W. Zhou et al., "Ensembled deep learning model outperforms human experts in diagnosing biliary atresia from sonographic gallbladder images," Nat Commun, vol. 12, no. 1, Art. no. 1, Feb. 2021, doi: 10.1038/s41467-021-21466-z.

[17] A. Esteva et al., "Dermatologist-level classification of skin cancer with deep neural networks," Nature, vol. 542, no. 7639, Art. no. 7639, Feb. 2017, doi: 10.1038/nature21056.

[18] E. P. Duhaime et al., "Nonexpert Crowds Outperform Expert Individuals in Diagnostic Accuracy on a Skin Lesion Diagnosis Task," Apr. 2023, doi: https://doi.org/10.1109/isbi53787.2023.10230646.

[19] L. Hong and S. E. Page, "Groups of diverse problem solvers can outperform groups of high-ability problem solvers," Proceedings of the National Academy of Sciences, vol. 101, no. 46, pp. 16385–16389, Nov. 2004, doi: 10.1073/pnas.0403723101.